\documentclass[12pt]{article}
\pdfoutput=1
\usepackage[utf8]{inputenc}
\usepackage[DIV=13]{typearea}
\usepackage{ragged2e}
\usepackage{amsmath,amsfonts,bbm,mathrsfs,amssymb}
\usepackage{slashed,cancel}
\usepackage{cite}
\usepackage{bm} 
\usepackage{hyperref}
\hypersetup{colorlinks=true,urlcolor=magenta,anchorcolor=blue,citecolor=blue,filecolor=blue,
            linkcolor=magenta,menucolor=blue, linktocpage=true,pdfproducer=medialab}
\usepackage[english]{babel}
\usepackage{catchfile}
\usepackage{indentfirst}
\usepackage{graphicx}
\usepackage{float}
\usepackage{enumerate}
\usepackage{caption}
\usepackage{subfigure}
\usepackage{multirow}
\usepackage[normalem]{ulem} 
\usepackage{booktabs}
\usepackage{physics}
\usepackage{bbold}
\usepackage{comment}
\usepackage{soul}
\newcommand{\email}[1]{\href{mailto:#1}{\tt #1}}

\numberwithin{equation}{section}

\newcommand{\magenta}[1]{\color{magenta} #1 \color{black}}
\newcommand{\be}{\begin{equation}}
\newcommand{\ee}{\end{equation}}
\newcommand{\ba} {\begin{equation}\begin{aligned}}
\newcommand{\ea} {\end{aligned}\end{equation}}
\newcommand{\sw}{\ensuremath{s_w}}

\newcommand{\tpdf}[2]{\texorpdfstring{#1}{#2}}

\renewcommand{\a}{\alpha}

\newcommand{\g}{\gamma}

\renewcommand{\to}{\rightarrow}

\newcommand{\Lag}{\mathscr{L}}
\newcommand{\de}{\partial}

\newcommand{\GeV}{\ \text{GeV}}

\newcommand{\ce}{\mathbf{c}_E}
\newcommand{\cL}{\mathbf{c}_L}

\newcommand{\ME}{\mathbf{M}_{E}}
\newcommand{\Mn}{\mathbf{M}_{\nu}}
\newcommand{\Cee}{\mathbf{c}_{ee}}
\newcommand{\Cnn}{\mathbf{c}_{\nu\nu}}

\def\Tr{{\rm Tr}}

\newcommand\subsetsim{\mathrel{%
  \ooalign{\raise0.2ex\hbox{$\subset$}\cr\hidewidth\raise-0.8ex\hbox{\scalebox{0.9}{$\sim$}}\hidewidth\cr}}}

\def\Tr{{\rm Tr}}

\newcommand{\cW}{c_{\widetilde W}}
\newcommand{\cB}{c_{\widetilde B}}
\newcommand{\cG}{c_{\widetilde G}}

\begin{document} 
\renewcommand*{\thefootnote}{\fnsymbol{footnote}}
\begin{titlepage}

\vspace*{-1cm}
\flushleft{\magenta{IFT-UAM/CSIC-23-121}}
\\[1cm]

\begin{center}
\bf\LARGE Limits on ALP-neutrino couplings from loop-level processes
\centering
\vskip .3cm
\end{center}
\vskip 0.5  cm
\begin{center}
{\large\bf J.~Bonilla}~\footnote{\email{jesus.bonilla@uam.es}},
{\large\bf B.~Gavela}~\footnote{\email{belen.gavela@uam.es}},
{\large\bf J.~Machado-Rodr\'iguez}~\footnote{\email{jonathan.machado@uam.es}},\\[2mm]
\vskip .7cm
{\footnotesize
Departamento de F\'isica Te\'orica and Instituto de F\'isica Te\'orica UAM/CSIC,\\
Universidad Aut\'onoma de Madrid, Cantoblanco, 28049, Madrid, Spain. \\
Instituto de F\'isica Te\'orica IFT-UAM/CSIC, Cantoblanco, E-28049, Madrid, Spain.
}
\end{center}
\vskip 2cm
\begin{abstract}
\justify
We obtain new bounds on effective ALP-neutrino interactions from their loop-level impact on the ALP couplings to electroweak gauge bosons $\g Z$, $ZZ$ and $W^+W^-$. Both types of interactions are connected via the chiral anomaly,  which allows to overcome the fermion mass suppression of tree-level analyses.  ALP couplings to charged leptons are considered simultaneously with ALP-neutrino ones, as required by gauge invariance. Complete one-loop computations are presented together with relevant kinematic limits.  We leverage data obtained from rare meson decay measurements alongside observations from collider experiments, among others. The novel constraints are particularly competitive for  ALP masses $\gtrsim 100$ MeV,  evidencing the powerful potential of future ALP searches at colliders which target its couplings to electroweak gauge bosons.
\end{abstract}
\end{titlepage}
\setcounter{footnote}{0}

\pdfbookmark[1]{Table of Contents}{tableofcontents}
\tableofcontents

\renewcommand*{\thefootnote}{\arabic{footnote}}

\newpage
\section{Introduction}

The quest for axions that solve the strong CP-problem, and more generally ALPs (axion-like particles), is ongoing with increasing intensity. Their interest stems from their nature as pseudo-Goldstone bosons (pGBs) which are singlets of the Standard Model of Particle Physics (SM), and are thus the generic tell-tale of yet undiscovered symmetries in Nature, classically almost exact but hidden (aka spontaneously broken). Beyond axions, many extensions of the SM predict pGBs which are SM singlets, such as for instance  the \emph{Majoron} associated to a dynamical nature of neutrino masses \cite{Gelmini:1980re,Langacker:1986rj,Ballesteros:2016xej,Arias-Aragon:2020qip,Arias-Aragon:2022ats},   pGBs from flavour symmetries known as {\it flavons} \cite{Davidson:1981zd,Wilczek:1982rv,Calibbi:2016hwq,Ema:2016ops},  or within composite Higgs models \cite{Georgi:1984af}, extra-dimensional theories \cite{Cicoli:2013ana,Bellazzini:2017neg},  and phenomenological string models with their plethora of spontaneously broken $U(1)$ global symmetries, as well as pGBs in  many inflationary theories \cite{Damour:1994zq}.

Because ALPs are pGBs, their new physics scale  $f_a$ is much larger than their mass $m_a\ll f_a$.  They are thus conveniently studied within the model-independent framework of Effective Field Theory (EFT), i.e. in terms of  a tower of operators weighted down by inverse powers of $f_a$. The leading operators have mass dimension five:  purely derivative couplings as pertains GBs, plus shift-breaking anomalous couplings and a mass term for the ALP. In analyses of generic ALPs --to be considered below-- the parameters $\{m_a, f_a\}$ are treated as free and independent (unlike in true-axion theories which solve the strong CP problem, for which their product is dynamically fixed).
 
 For ALPs, the range of possible values of $f_a$ and $m_a$ is wide, leading to an incredibly rich phenomenological arena for particle and astroparticle physics.  To compound their intrinsic interest, both axions and ALPs can be excellent candidates to explain the nature of dark matter (DM). An intense theoretical exploration is ongoing on ALP EFTs and their predictions for laboratory experiments, astrophysics and cosmology, together with the construction of ultraviolet (UV) complete theories.  In synergy with it, a plethora of experiments, which look  for ALPs in different regimes and with diversified techniques, are either already taking data or scheduled to do so in the next decade~\cite{Choi:2020rgn}.

We address in this work the effective interactions between ALPs and neutrinos.  At leading order in the ALP EFT these appear as combinations of ALP derivative insertions $\partial_\mu a$ and fermionic currents, that is,  in the form of chirality-conserving fermion bilinears with arbitrary coefficients. Their  tree-level impact on observables turns out to be proportional to the fermion masses,  as it can be straightly inferred from the classical equations of motion (EOM):   they show that --at the classical level-- the  ALP-fermionic operators are tradable by chirality-flipping ones proportional to the fermion masses~\cite{Chala:2020wvs,Bauer:2020jbp,Bonilla:2021ufe,Bonilla:2022qgm}. Given the tiny masses of neutrinos, their tree-level contribution is  thus extremely suppressed and negligible.  In contrast, this mass suppression is overcome when the system is contemplated including one-loop effects. This is obvious noting that the use of EOMs is tantamount to a chiral rotation which necessarily induces at one-loop anomalous ALP couplings, and in consequence  mass-independent components.  The present experimental constraints on ALP couplings to $\g Z$, $ZZ$ and $W^+W^-$ will be shown to be strong enough to suggest relevant bounds on ALP-neutrino interactions, in spite of the $\mathcal{O} (\alpha/ \pi)$ weight of one-loop effects.  The loop-induced impact of ALP-neutrino interactions on the coupling of ALPs to electroweak (EW)  bosons will be shown to probe regions of the parameter space for ALP-lepton couplings that are otherwise unexplored.
 
 The ALP EFT theory is constructed for scales  $f_a\gg v\simeq 246 \GeV$ where $v$ is the SM electroweak (EW) scale. A proper EFT formulation must then be explicitly $SU(3)\times SU(2) \times U(1)$ gauge invariant above the electroweak scale. This brings into the game the couplings of ALPs to charged leptons, as the left-handed components of the latter unavoidably accompany  left-handed neutrinos. Bounds on charged ALP-lepton couplings  have been previously explored~\cite{ciaran_o_hare_2020_3932430,Caputo:2021rux,Bauer:2021mvw,Ferreira:2022xlw,Oikonomou:2023qfl} and these will be taken into account in our analysis. Furthermore, recent works\cite{Huang:2018cwo,Reynoso:2022vrn,Lichkunov:2023iyt} propose  to study the impact of DM ALPs on neutrino propagation and oscillations, and cosmology, but they explore a much lighter range of ALP masses than that considered in this paper.

We present the complete one-loop corrections to the couplings of ALPs to  on-shell EW gauge boson pairs, $\g\g$, $\g Z$, $ZZ$ and $WW$, induced by ALP couplings to  neutrinos and charged leptons. Our work expands and completes previous corrections involving ALP fermionic currents~\cite{Bauer:2017ris,Chala:2020wvs,Bauer:2020jbp,Bonilla:2021ufe,Ferreira:2022xlw}. Neutrino masses will be disregarded, unless otherwise stated in which case the role of PMNS mixing will be discussed. Various kinematics regions and limits are contemplated: the best --and novel-- bounds on ALP-neutrino interactions will turn out to be those for relatively heavy ALPs, $m_a \gtrsim 100$ MeV up to 1 TeV, correlated with the fact that the strongest constraints on ALP  couplings to heavy EW bosons stem from collider searches and from rare meson decay searches.
 
An additional comment is pertinent in the framework of the most general ALP EFT.  The latter may include from the start dimension five anomalous ALP couplings to gauge bosons, with arbitrary coefficients. In consequence, the contribution stemming from the chiral rotation described above will confront experiment mixed with that from the putative anomalous gauge couplings of ALPs in the initial Lagrangian. In spite of this,   bounds on the coefficients of ALP-neutrino flavour-diagonal couplings can be extracted from experiment,  barring  fine-tuned cancellations among the contributions from both sources.

The structure of this manuscript can be inferred from the Table of Contents.

\section{The ALP Effective Lagrangian}\label{sec:Lagrangian}

The ALP EFT Lagrangian,  up to operators with mass dimension five, reads~\cite{Georgi:1986df,Choi:1986zw,Bonilla:2021ufe,Bonilla:2022qgm}
\begin{equation}
\label{eq:L_ALP}
\Lag_{\rm ALP} = \frac{1}{4}\de_\mu a\de^\mu a - \frac{1}{2} m_a^2 a^2 + \Lag_a^{\rm gauge}  + \Lag_{\de a}^{\rm fermion} \,,
\end{equation}
where $\Lag_a^{\rm gauge}$ stands for the anomalous ALP-gauge boson effective couplings, which at energies above EW symmetry breaking can be written as
\begin{equation} \label{eq:L_ALP_boson}
    \Lag_a^{\rm gauge} =  - \cG \frac{a}{f_a} G_{\mu \nu}^\a \widetilde{G}^{\mu \nu,\, \a}  - \cW \frac{a}{f_a} W_{\mu \nu}^i \widetilde{W}^{\mu \nu,\, i}  - \cB \frac{a}{f_a} B_{\mu \nu} \widetilde{B}^{\mu \nu} \,,
\end{equation}
with $G_{\mu \nu}^\a$, $W_{\mu \nu}^i$ and $B_{\mu \nu}$ denoting the field strength of the strong and EW gauge bosons respectively ($i = 1, 2, 3$ and $\a = 1, ... , 8$), while $\Lag_{\de a}^{\rm fermion}$ stands for the gauge-invariant ALP-fermion derivative interactions, 
\begin{equation} \label{eq:L_ALP_fermion}
    \Lag_{\de a}^{\rm fermion} = \sum_{\Psi} \frac{\de_\mu a}{f_a} \, \overline{\Psi} \gamma^\mu \mathbf{c}_\Psi \Psi \,,
\end{equation}
where  the sum runs over the chiral fermionic fields of the SM: $\Psi = \left\{ Q_L, L_L, u_R, d_R, e_R \right\}$, each $\Psi$ being a 3-component vector in flavour space and  
 $\mathbf{c}_\Psi$   $3\times 3$ being hermitian matrices in that space.\footnote{Note that there exists in addition a   shift-invariant bosonic operator involving the Higgs doublet, $\mathcal{O}_{a\Phi}=\partial_\mu a (\Phi^\dag i\overleftrightarrow{D_\mu}\Phi)/{f_a}$, which has {\it not} been included in Eq.~\eqref{eq:L_ALP}, as it would be redundant given the choice made to consider all possible fermionic couplings in $\Lag_{\partial a}^\psi$~\cite{Bonilla:2021ufe}.} Note that no coupling to putative right-handed neutrinos is considered, as  neutrino masses will be neglected below. In all generality, those  chiral fields in the $\Psi$ set  are not the left and right components of fermion mass eigenstates. 
   Nevertheless, the rotation  of the $\mathbf{c}_\Psi$ matrices to the mass basis~\cite{Chala:2020wvs,Bauer:2020jbp,Bonilla:2021ufe,Bonilla:2022qgm} is not relevant for the main results of this paper, which   focuses on purely leptonic couplings neglecting neutrino masses (i.e. all $\mathbf{c}_\Psi$ matrices will be disregarded except for $\ce$ and $\cL$). $\Lag_{\partial a}^\psi$ simplifies then to
\begin{equation} \label{eq:L_ALP_lepton}
\begin{aligned}
    \Lag_{\de a}^{\rm fermion} \, & = \frac{\de_\mu a}{f_a} \, \overline{L}_L \gamma^\mu \cL L_L + \frac{\de_\mu a}{f_a} \, \overline{e}_R \gamma^\mu \ce e_R \,, \\
    & = \frac{\de_\mu a}{f_a} \, \overline{\nu}_L \gamma^\mu \cL \nu_L + \frac{\de_\mu a}{f_a} \, \overline{e} \gamma^\mu \left(\ce P_R + \cL P_L \right) e \,, \\
    & = \frac{\de_\mu a}{2 f_a} \, \overline{\nu}_L \gamma^\mu (1 - \g^5) \cL \nu_L + \frac{\de_\mu a}{2 f_a} \, \overline{e} \gamma^\mu \left(\left( \ce + \cL \right) + \left( \ce - \cL \right) \g^5 \right) e \,,
\end{aligned}
\end{equation}
where $P_{R,L}$ stands for the  chirality projectors, and in the last two lines the left-handed EW lepton fields have been decomposed on their electrically charged and neutral components $L_L=(e_L, \nu_L)$, with $e= e_R + e_L$. This  illustrates that the ALP-neutrino couplings are defined only in terms of $\cL$, while their analysis  
 unavoidably requires to contemplate as well the couplings of charged leptons to ALPs.
 
 It is worth to remind that  in flavour-diagonal scenarios {\it and   at tree-level}, the ``phenomenological'' couplings to the physical leptons only depend  on the axial component of the operators in Eq.~\eqref{eq:L_ALP_lepton} (see Ref.~\cite{Bonilla:2021ufe,Bonilla:2022qgm}),
 \begin{equation}
    \Lag^{\rm fermion}_{\partial a} \supset \frac{\partial_\mu a}{f_a} \sum_{\ell} \left(\Cnn \right)_{\ell\ell} \overline{\nu}_\ell \gamma^\mu \gamma^5 \nu_\ell + \frac{\partial_\mu a}{f_a} \sum_{\ell} \left(\Cee \right)_{\ell\ell} \overline{\ell} \gamma^\mu \gamma^5 \ell \,,
\end{equation}
 where $\Cee$ and $\Cnn$ are defined as
\begin{equation} \label{eq:pheno-couplings}
    \Cee \equiv (\ce - \cL) \,, \qquad \Cnn \equiv (- \cL) \,.
\end{equation}
   Indeed, the orthogonal combination, i.e. the vectorial coupling, vanishes due to the conservation of vectorial currents (lepton number) at classical level.\footnote{This does not hold for flavour non-diagonal couplings, which  can be relevant in practice when considering ALP decays to charged leptons (e.g. ALP$\to e\mu$); these will not be considered in this work.} At the quantum level, though, lepton number currents are broken due to chiral anomalies and the vector components do contribute even in the flavour-diagonal case: they will induce interactions between the ALP and heavy gauge bosons,  which will be taken into account. In any case, because in all generality there are only two independent set of couplings, $\{ \ce, \cL\}$, we choose to show our results below in terms of the  two phenomenological combinations defined in Eq.~(\ref{eq:pheno-couplings}), as they are more transparent to confront experimental data.
   
At energies below EW symmetry breaking, the bosonic Lagrangian $\Lag_a^{\rm gauge}$ in Eq.~\eqref{eq:L_ALP_boson} can be rewritten  in terms of the EW gauge boson mass eigenstates,
\begin{equation} \label{eq:L_ALP_boson_physical}
\begin{aligned}
    \Lag_{a}^{\rm gauge} =\, & -\frac{1}{4} g_{agg} \,a\, G_{\mu \nu}^\a \widetilde{G}^{\mu \nu,\, \a}  - \frac{1}{4} g_{a\g\g} \,a\,F_{\mu \nu} \widetilde{F}^{\mu \nu} \\
    & - \frac{1}{4} g_{a\g Z} \,a\,F_{\mu \nu} \widetilde{Z}^{\mu \nu} - \frac{1}{4} g_{aZZ} \,a\,Z_{\mu \nu} \widetilde{Z}^{\mu \nu} - \frac{1}{2} g_{aWW} \,a\,W_{\mu \nu}^+ \widetilde{W}^{\mu \nu,\,-} \,,
\end{aligned}
\end{equation}
with
\begin{equation} \label{eq:EW_couplings}
    \begin{aligned}
        & g_{agg} \equiv \frac{4}{f_a} \cG \,, \qquad & g_{a\g\g}  \equiv \frac{4}{f_a} \left( c_w^2 \cB + s_w^2 \cW \right) \,,  \qquad & \\
        & g_{aWW}   \equiv \frac{4}{f_a} \cW \,, \qquad & g_{aZZ}  \equiv \frac{4}{f_a} \left( s_w^2 \cB + c_w^2 \cW \right) \,,  \qquad & g_{a\g Z}  \equiv \frac{8}{f_a} c_w s_w \left( \cW - \cB \right) \,,
    \end{aligned}
\end{equation}
where $c_w (s_w)$ is the cosine (sine) of the weak mixing angle.

\subsubsection*{Leptonic EOM}
   The low-energy equations of motion (EOM) for the lepton fields, taking into account one-loop anomalous corrections, can be written  as
    \begin{equation}
    \begin{aligned}
    \frac{\partial_\mu a}{f_a}\Bar{e}_R \gamma^\mu \ce e_R =&  -\left( i\frac{a}{f_a} \Bar{e}_L \ME \ce e_R  + \text{h.c.} \right) + \text{Tr}\left[\ce \right] \frac{a}{f_a} \frac{g'^2}{16 \pi^2} B_{\mu\nu}\Tilde{B}^{\mu\nu}\,, \\
    \frac{\partial_\mu a}{f_a}\Bar{e}_L \gamma^\mu \cL e_L=&  \left( i\frac{a}{f_a} \Bar{e}_L \ME \cL e_R  + \text{h.c.} \right) - \text{Tr}\left[\cL \right] \frac{a}{f_a} \Bigg[ \frac{g'^2}{64 \pi^2} B_{\mu\nu}\Tilde{B}^{\mu\nu} + \frac{g^2}{64 \pi^2} W_{\mu\nu}\Tilde{W}^{\mu\nu} \Bigg] \,, \\
    \frac{\partial_\mu a}{f_a}\Bar{\nu}_L \gamma^\mu \cL \nu_L=&  \left( i\frac{a}{f_a} \Bar{\nu}_L \Mn \cL \nu_R  + \text{h.c.} \right) - \text{Tr}\left[\cL \right] \frac{a}{f_a} \Bigg[ \frac{g'^2}{64 \pi^2} B_{\mu\nu}\Tilde{B}^{\mu\nu} + \frac{g^2}{64 \pi^2} W_{\mu\nu}\Tilde{W}^{\mu\nu} \Bigg] \,, \label{ultima}
    \end{aligned}
    \end{equation}
    where $\ME $ and $\Mn$ denote here the generic mass matrices for charged leptons and neutrinos, respectively, and where the anomalous contributions have been left  expressed here  in terms of the hypercharge and $SU(2)_L$ field strengths, $B_{\mu\nu}$ and $W_{\mu\nu}$ respectively, for notation compactness. Given the tiny values of neutrinos masses, Eq.~(\ref{ultima}) already shows that experimental bounds on ALP couplings to massive EW gauge bosons directly suggest limits on the flavour-diagonal components of ALP-neutrino couplings, barring fine-tuned cancellations with tree-level gauge couplings.

\section{One-loop induced couplings}\label{sec:loops}

\begin{figure}[t]\centering
\begin{center}
	\includegraphics[width=0.35\textwidth]{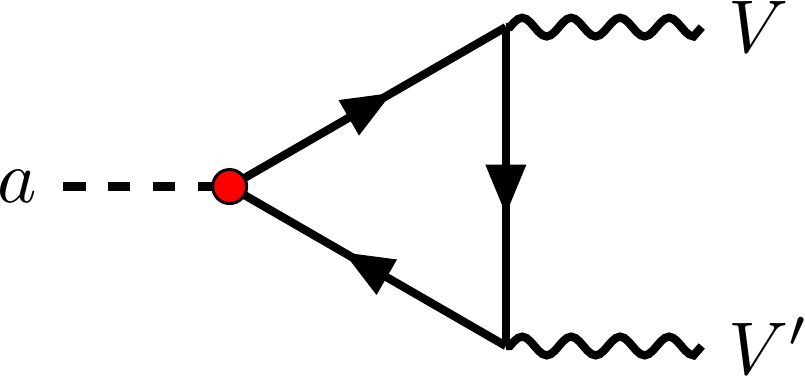}
\end{center}
\caption{One-loop diagram contributing to the ALP-gauge effective couplings originated from the ALP-fermion coupling.}
\label{figdiagram}
\end{figure}

In this section we present the one-loop contribution of ALP-lepton couplings to  the strength of  the ALP-EW gauge bosons $\{g_{a\g\g} , g_{aW W} , g_{aZZ} , g_{a\g Z} \}$ defined in Eq.~\eqref{eq:EW_couplings}. These corrections correspond to the triangle diagram in Fig.~\ref{figdiagram}.  For completeness, both the anomalous and mass-dependent one-loop corrections will be developed. For the computation of the latter, the ALP field $a$ and the SM bosons will be taken to be on-shell, and thus the amplitudes will be written in terms of the gauge boson masses $M_W$ and $M_Z$, the charged lepton masses $m_\ell$ and the ALP mass $m_a$. Nevertheless, the amplitude for any coupling with the ALP off-shell can be immediately  obtained from the expressions below by simply replacing $m_a^2 \to p_a^2$, where $p_a$ is the  ALP four-momentum. All computations have been performed with the help of the {\tt Mathematica} packages {\tt FeynCalc} and {\tt Package-X}~\cite{Patel:2016fam,Shtabovenko:2020gxv}.  
 
Although neutrino masses are to be disregarded in the main text because they induce negligible corrections, their impact is nevertheless made explicit  for the no flavour-mixing case in  App.~\ref{app:full-expressions}.   In turn, lepton-flavour mixing effects would be {\it a priori} present for final state $W$ bosons, but the amplitudes would remain in practice equivalent to those for the no flavour-mixing scenario due to the unitarity of the PMNS matrix: we have verified that the GIM-suppressed flavour corrections are of order $\mathcal{O} (m_\nu^2 m_\ell^2 / M_W^4)$ and thus totally negligible.

Without further ado, we present next the results for the amplitudes computed from the triangle diagram in Fig.~\ref{figdiagram}, in the limit $m_\nu \to 0$ for all neutrino flavours.

\paragraph{Contributions to \boldmath $g_{a\g\g}$\unboldmath:} 
Previous computations of the ALP-photon interaction induced at one-loop by ALP-charged lepton couplings can be found in Refs.~\cite{Bauer:2017ris,Bauer:2020jbp,Bonilla:2021ufe}, whose results we verified. They stem from the diagram in Fig.~\ref{figdiagram} with charged leptons running in the loop. Being electrically neutral, neutrinos do not play a role in this two-photon channel, and the induced coupling reads
\begin{equation} \label{eq:photonphoton}
\begin{aligned}
    g_{a\g\g}^{\rm loop} = \, & - \frac{\alpha_{\rm em}}{\pi f_a} \left[ \Tr{(\Cee)}  + 2 \sum_{\ell  } (\Cee)_{\ell\ell} m_{\ell}^2 \mathcal{C}_0 \left( 0, 0, m_a^2, m_{\ell}, m_{\ell} , m_{\ell} \right)  \right] \\
    = & - \frac{\alpha_{\rm em}}{\pi f_a} \left[ \Tr{(\Cee)}  - \sum_\ell (\Cee)_{\ell\ell} \tau_\ell f \left( \tau_\ell \right)^2 \right] \,,
\end{aligned}
\end{equation}
where $\ell = e, \mu, \tau$ here and all through the paper,  $\tau_\ell \equiv 4 m_{\ell}^2 / m_a^2$, and $\mathcal{C}_0$ is the scalar 3-point Passarino-Veltman function (see Ref.~\cite{Passarino:1978jh}), 
\begin{equation}
\begin{aligned}\label{eq:function_C0}
& \mathcal{C}_0 (q_1^2 , q_2^2 , p^2 , m_1 , m_2 , m_3) \equiv \\
& \int_0^1 \text{d}x \int_0^x \frac{\text{d}y }{(x-y)y q_1^2 - (x-y) (x-1) q_2^2 -y (x-1) p^2 -y m_1^2 - (x-y) m_2^2 + (x-1) m_3^2} \,,
 \end{aligned} 
\end{equation}
and 
\begin{equation}
    f (\tau) =  \begin{cases} \arcsin \frac{1}{\sqrt{\tau}} & \mbox{for } \tau \ge 1 \,, \\ \frac{\pi}{2} + \frac{i}{2} \log{\frac{1 + \sqrt{1 - \tau}}{1- \sqrt{1 - \tau}}} & \mbox{for } \tau < 1 \,. \end{cases}
    \label{eq:f_tau}
\end{equation}
This shows that the size of the effective ALP-photon coupling in Eq.~\eqref{eq:photonphoton} only depends on the ratio $ m_{\ell}^2 /m_a^2$  (assuming that the ALP and both photons are on-shell particles). In particular, in the heavy ALP limit, $m_a \gg m_{\ell}$, the second term in Eq.~\eqref{eq:photonphoton}, which is mass-dependent, becomes negligible, and then only the first term remains:
\begin{equation}
    g_{a\g\g}^{\rm loop} \approx - \frac{\alpha_{\rm em}}{\pi f_a} \Tr (\Cee) \qquad \mbox{ for } m_{\ell} \ll m_a \,.
\end{equation}
This first term is the \emph{anomalous} term of the triangle diagram, which is mass-independent. Basically this term corresponds to the chiral anomaly from the fermion current to which the ALP is coupled in Eq.~\eqref{eq:L_ALP_lepton}. On the other hand, in the light ALP limit $m_a \ll m_{\ell}$, both terms --the anomalous and the mass-dependent one-- partially cancel each other and the net  coupling  is suppressed as 
\begin{equation}\label{suppressed-gphoton}
    g_{a\g\g}^{\rm loop} \approx - \frac{\alpha_{\rm em}}{12 \pi f_a} \sum_\ell (\Cee)_{\ell\ell} \frac{m_a^2}{m_{\ell}^2} \qquad \mbox{ for } m_{\ell} \gg m_a \,. 
\end{equation}
This is a well known property of the triangle diagram for vector-like gauge interactions and massless gauge bosons such as  QED~\cite{Quevillon:2019zrd}. This behaviour will not hold in the presence of  massive gauge bosons, i.e. the $Z$ and $W$ bosons, see below.

\paragraph{Contributions to \boldmath $g_{a\g Z}$\unboldmath:}  Even though neutrinos --being electrically neutral-- cannot run in the loop that induces $g_{a\gamma Z}$, this ALP-gauge bosons coupling can still be used to measure the physical couplings between ALPs and neutrinos as shown by its dependence on  $\Cnn$. Indeed, the one-loop-induced ALP-$\g Z$ coupling reads:
\begin{equation} \label{eq:photonZ}
\begin{aligned}
    g_{a\g Z}^{\rm loop} \, & = \frac{\alpha_{\rm em}}{c_w s_w \pi f_a} \left[ \Tr{(\Cnn)} - 2 s_w^2 \Tr{(\Cee)}   + (1 - 4 s_w^2) \sum_\ell (\Cee)_{\ell\ell} m_{\ell}^2 \mathcal{C}_0 \left( 0, M_Z^2, m_a^2, m_{\ell}, m_{\ell} , m_{\ell}\right) \right] \,, \\
    & = \frac{\alpha_{\rm em}}{c_w s_w \pi f_a} \left[ \Tr{(\Cnn)} - 2 s_w^2 \Tr{(\Cee)}  - 2(1 - 4 s_w^2) \sum_\ell  \frac{(\Cee)_{\ell\ell} m_{\ell}^2}{m_a^2 - M_Z^2} \Big( f(\tau_\ell)^2 - f(\tau_\ell^Z)^2\Big) \right] \,,
\end{aligned}
\end{equation}
where $\tau_\ell^Z \equiv 4 m_{\ell}^2 / M_Z^2$ (previous computations can be found in Refs.~\cite{Bauer:2020jbp,Bonilla:2021ufe}).  The  dependence on $\Cnn$ is a consequence of  the chiral anomaly since the $Z$ boson couples with different weak charges to right-handed and left-handed leptons, and thus  the equation exhibits the $\cL$ contribution which is shared with charged leptons, see Eq.~(\ref{eq:pheno-couplings}).

Once again two distinct types of contributions appear:  the first two terms in Eq.~(\ref{eq:photonZ}) are anomalous ones and thus mass-independent, while the mass-dependent term is proportional to $m_{\ell}^2$ times the contribution from  $f(\tau)$ functions. The latter is induced exclusively by the axial combination of couplings. Given  that the $Z$ boson is much heavier than any SM lepton, the following approximation for the $f(\tau^Z)$ function in Eq.~\eqref{eq:photonZ} is pertinent:
\begin{equation}
    f(\tau_\ell^Z)^2 \approx \frac{1}{4} \left[ \pi + i \log \left( \frac{M_Z^2}{m_{\ell}^2} \right) \right]^2 \,.
\end{equation}
It follows that the mass-dependent term in Eq.~(\ref{eq:photonZ}) is always suppressed, at least, by a factor $m_\ell^2/m_Z^2$, 
in contrast with the case for $g_{a\g\g}$ in Eq.~(\ref{eq:photonphoton}). This means that the anomalous term in $g_{a\g Z}$ will not be cancelled in any kinematic region of $m_a$. In other words, the anomalous term will always dominates $g_{a\g Z}$ while the mass-dependent term provides at most a correction of order $\mathcal{O} (m_{\ell}^2 / M_Z^2)$,
\begin{equation}
    g_{a\g Z}^{\rm loop} \approx\frac{\alpha_{\rm em}}{c_w s_w \pi f_a} \left[ \Tr{(\Cnn)} - 2 s_w^2 \Tr{(\Cee)}  + \mathcal{O} \left( \frac{m_{\ell}^2}{M_Z^2} \right) \right] \,.
\end{equation}
This property will also hold for all ALP couplings to heavy EW bosons considered below.

Simplified formulae of interest for different regimes of ALP mass versus $m_{\ell}$ and $M_Z$ read:

\begin{itemize}
\item For $m_a \ll m_{\ell} \ll M_Z$:
\end{itemize}
\begin{equation}
\begin{aligned}
    g_{a\g Z}^{\rm loop} \approx \frac{\alpha_{\rm em}}{c_w s_w \pi f_a} \left\{ \Tr{(\Cnn)} - 2 s_w^2 \Tr{(\Cee)}   - (1 - 4 s_w^2)\sum_\ell   \frac{(\Cee)_{\ell\ell}  m_{\ell}^2}{2 M_Z^2} \left[ \pi + i \log \left( \frac{M_Z^2}{m_{\ell}^2} \right) \right]^2 \right\} \,.
\end{aligned}
\end{equation}
\begin{itemize}
\item For $m_{\ell} \ll m_a \ll M_Z$:
\end{itemize}
\begin{equation}
\begin{aligned}
    g_{a\g Z}^{\rm loop} \approx \frac{\alpha_{\rm em}}{c_w s_w \pi f_a} \, & \Bigg\{ \Tr{(\Cnn)} - 2 s_w^2 \Tr{(\Cee)} \\
    & \left. - (1 - 4 s_w^2)\sum_\ell  \frac{(\Cee)_{\ell\ell} m_{\ell}^2}{M_Z^2} \log\left(  \frac{m_a^2 M_Z^2}{m_\ell^4}\right) \left[ \log \left( \frac{m_a}{M_Z} \right) + i \pi \right] \right\} \,.
\end{aligned}
\end{equation}

\begin{itemize}
\item For $m_{\ell} \ll M_Z \ll m_a$:
\end{itemize}
\begin{equation}
\begin{aligned}
    g_{a\g Z}^{\rm loop} \approx \frac{\alpha_{\rm em}}{c_w s_w \pi f_a} \, & \Bigg\{ \Tr{(\Cnn)} - 2 s_w^2 \Tr{(\Cee)} \\
    & \left. -(1 - 4 s_w^2) \sum_\ell  \frac{(\Cee)_{\ell\ell}  m_{\ell}^2}{m_a^2} \log\left(  \frac{m_a^2 M_Z^2}{m_\ell^4}\right) \left[ \log \left( \frac{M_Z}{m_a} \right) + i \pi \right] \right\} \,.
\end{aligned}
\end{equation}

\paragraph{Contributions to \boldmath $g_{aZZ}$\unboldmath:} The complete  correction to the on-shell $a$-$ZZ$ coupling induced  at one-loop by ALP-neutrino and ALP-electron couplings reads
\begin{equation} \label{eq:ZZ_1_family}
    \begin{aligned}
        g_{aZZ}^{\rm loop} =  \, & - \frac{\alpha_{\rm em}}{2 c_w^2 s_w^2 \pi f_a} \Bigg\{ ( 1 - 2 s_w^2 ) \Tr (\Cnn)  + 2 s_w^4 \Tr (\Cee)  \\
        & - \sum_\ell \frac{(\Cee)_{\ell\ell} m_{\ell}^2}{m_a^2 - 4 M_Z^2} \Bigg[ \mathcal{B} (m_a^2, m_{\ell}, m_{\ell}) - \mathcal{B} (M_Z^2, m_{\ell}, m_{\ell}) \\
        & + \left[ (1 - 4 s_w^2)^2 M_Z^2  + 2 s_w^2 (1 - 2 s_w^2) m_a^2  \right] \mathcal{C}_0 \left( m_a^2, M_Z^2, M_Z^2, m_{\ell}, m_{\ell} , m_{\ell}\right) \Bigg] \Bigg\} \,,
    \end{aligned}
\end{equation}
where $\mathcal{B}$ is the function \texttt{DiscB} in \texttt{Package-X} (see Ref.~\cite{Bonilla:2021ufe}), which can be written as 
\begin{equation} \label{functionDB}
\mathcal{B} (p^2 , m_1 , m_2 ) \equiv \frac{ \sqrt{\lambda(p^2,m_1^2,m_2^2)}}{p^2} \log \left( \frac{m_1^2 + m_2^2 - p^2 + \sqrt{\lambda(p^2,m_1^2,m_2^2)}}{2 m_1 m_2} \right) \,,
\end{equation}
in which $\lambda$ is the K\"{a}ll\'en triangle function: $\lambda (a,b,c) \equiv a^2 + b^2 + c^2 - 2 ab - 2 bc - 2 ca$. As in the previous case, given that $m_\ell \ll M_Z$ for all SM leptons,  the 
$\mathcal{B}$ functions  in Eq.~\eqref{eq:ZZ_1_family} can be approximated as 
\begin{equation}
    \mathcal{B} (M_Z^2, m_{\ell}, m_{\ell}) \approx  i \pi + \log \left( \frac{m_{\ell}^2}{M_Z^2} \right) \,,
\end{equation}
and 
\begin{equation}
    \mathcal{C}_0 \left( m_a^2, M_Z^2, M_Z^2, m_{\ell}, m_{\ell} , m_{\ell}\right) \approx \mathcal{C}_0 \left( m_a^2, M_Z^2, M_Z^2, 0,0,0 \right)\,.
\end{equation}

In the assumption that all particles are  on-shell,   the only physical process stemming from the $a$-$ZZ$ vertex in Fig.~\ref{figdiagram} corresponds to the decay of a heavy ALP into two $Z$ bosons, for  which the relevant kinematic regime is    $m_\ell \ll m_a$, and thus
\begin{equation} 
    \begin{aligned}
        g_{aZZ}^{\rm loop} \approx  - \frac{\alpha_{\rm em}}{2 c_w^2 s_w^2 \pi f_a} \, & \Bigg\{ ( 1 - 2 s_w^2 ) \Tr (\Cnn)  + 2 s_w^4 \Tr (\Cee) -  \sum_\ell  \frac{(\Cee)_{\ell\ell} m_{\ell}^2}{m_a^2 - 4 M_Z^2} \Bigg[ \log \left( \frac{M_Z^2}{m_a^2} \right) \\
        & + \Big( (1 - 4 s_w^2)^2 M_Z^2  + 2 s_w^2 (1 - 2 s_w^2) m_a^2  \Big) \mathcal{C}_0 \left(m_a^2, M_Z^2, M_Z^2, 0,0,0\right) \Bigg] \Bigg\} \,.
    \end{aligned}
\end{equation}
This expression can be simplified for ALPs much heavier than the $Z$ boson, 
\begin{equation}
\begin{aligned}
        g_{aZZ}^{\rm loop} \approx  - \frac{\alpha_{\rm em}}{2 c_w^2 s_w^2 \pi f_a} \, & \Bigg\{ ( 1 - 2 s_w^2 ) \Tr (\Cnn)  + 2 s_w^4 \Tr (\Cee) - 2\sum_\ell  \frac{(\Cee)_{\ell\ell} m_{\ell}^2}{m_a^2}\Bigg[ \log\left(\frac{M_Z}{m_a}\right) \\ & +i\pi +s_w^2 (1 - 2 s_w^2) \left(\frac{\pi^2}{3} + \log\left(\frac{M_Z^2}{m_a^2}\right)^2 \right)\Bigg] \Bigg\}\,.
    \end{aligned}
\end{equation}
Note that  this expression is also valid in the regime $m_{\ell} \ll M_Z \ll m_a$ for off-shell ALPs upon the replacement $m_a^2 \to p_a^2$ and as long as $M_Z^2 \ll p_a^2$.

\paragraph{Contributions to \boldmath $g_{aWW}$\unboldmath:} Finally, the correction to the $a$-$WW$ on-shell coupling induced at one-loop by the insertion of ALP-neutrino and ALP-electron couplings reads
\begin{equation} \label{eq:WW}
    \begin{aligned}
        g_{aWW}^{\rm loop} =  \, & -\frac{\alpha_{\rm em}}{2 s_w^2 \pi f_a} \Bigg\{ \Tr (\Cnn) -  \sum_\ell  \frac{(\Cee)_{\ell\ell} m_{\ell}^2}{m_a^2 - 4 M_W^2} \Bigg[ \mathcal{B} (m_a^2, m_{\ell}, m_{\ell}) \\
        & +\left( 1-\frac{m_{\ell}^2}{M_W^2}\right) \left( i \pi + \log \left( \frac{M_W^2}{m_{\ell}^2} -1\right) + M_W^2 \mathcal{C}_0 \left( m_a^2, M_W^2, M_W^2, m_{\ell}, m_{\ell} , 0\right) \right) \Bigg] \Bigg\} \,,
    \end{aligned}
\end{equation}
where the  second  (mass-dependent) term corresponds to the diagram in which the two fermionic legs attached to the ALP in Fig.~\ref{figdiagram} are electrons and the vertically exchanged lepton is a neutrino, and viceversa for the anomalous term. Note that all contributions are flavour diagonal at the leading order in which we work.\footnote{As an illustration of the  (in)dependence  with respects to the PMNS mixing matrix  $U^{\rm PMNS}$, it is easy to see that, if  leptonic mixing is considered, the  $(\Cee)_{\ell\ell}$ coupling  in the second term in Eq.~\eqref{eq:WW} would be simply replaced by  $ \sum_{i} U_{\nu_i \ell}^{\rm PMNS} (\Cee)_{\ell\ell'} (U_{\ell' \nu_i}^{\rm PMNS})^* = (\Cee)_{\ell\ell'} \delta_{\ell\ell'}$, recovering the original equation for the scenario with no lepton mixing.}

Taking again the limit $m_\ell \ll m_a , M_W$, the expression above simplifies to
\begin{equation} \label{eq:WW_small_ml}
\begin{aligned}
        g_{aWW}^{\rm loop} \approx  -\frac{\alpha_{\rm em}}{2 s_w^2 \pi f_a} \, & \Bigg\{ \Tr (\Cnn) -  \sum_\ell \frac{ (\Cee)_{\ell\ell} \,  m_{\ell}^2}{m_a^2 - 4 M_W^2} \Bigg[ \log \left( \frac{M_W^2}{m_a^2} \right)  \\
        & + M_W^2 \, \mathcal{C}_0 \left(m_a^2,  M_W^2, M_W^2, 0,0,0\right) \Bigg] \Bigg\} \,,
\end{aligned}
\end{equation}
which can be further approximated for an ALP much heavier than the $W$ boson,
    \begin{equation}
    \begin{aligned}
    \label{eq.gaWW-MWllma}
    g_{aWW}^\text{eff} \approx & -\frac{\alpha_{em}}{2 \sw^2 \pi f_a }\Bigg[ \Tr{(\Cnn)} - \sum_\ell \frac{ (\Cee)_{\ell\ell} \,  m_{\ell}^2}{m_a^2}\log \left( \frac{M_W^2}{m_a^2}\right) \Bigg] \,. 
    \end{aligned}
    \end{equation}


\section{Bounds on ALP-neutrino effective interactions}\label{sec:bounds}

We present next the bounds on ALP-neutrino coupling which follow from the experimental bounds on $\{g_{a\g\g} , g_{aW W} , g_{aZZ} , g_{a\g Z} \}$ couplings (barring fine-tuned cancellations with hypothetical tree-level anomalous couplings in Eq.~(\ref{eq:L_ALP_boson}), as explained earlier). We include bounds on the ALP-electron coupling as they impose further constraints on the ALP-neutrino sector, due to gauge invariance correlations between $\Cee$ and $\Cnn$ that arise from the coupling to left-handed leptons $\cL$, see Eq. (\ref{eq:pheno-couplings}). 

In the previous section it was shown that the anomalous terms in  the couplings of ALPs to massive gauge bosons suffice for the numerical analysis, given today's experimental precision, 
\begin{align}
 &
 g_{a\g Z}^{\rm loop} \approx - \frac{\a_{\rm em}}{s_w c_w \pi f_a}\left( 2 s_w^2 \Tr(\Cee)  - \Tr(\Cnn) \right) \,, \label{eq:photonZ_anomaly}
 \\
 &
 g_{aZZ}^{\rm loop} \approx - \frac{\a_{\rm em}}{2 s_w^2 c_w^2 \pi f_a}\left( 2 s_w^4 \Tr(\Cee) + \left( 1- 2 s_w^2 \right) \Tr(\Cnn) \right) \,, \label{eq:ZZ_anomaly}
 \\
 &
 g_{aWW}^{\rm loop} \approx - \frac{\a_{\rm em}}{2s_w^2 \pi f_a} \Tr (\Cnn) \,. \label{eq:WW-anomaly}
\end{align}
This is in contrast with the ALP-photon channel in the regime of light ALPs, $m_a  \lesssim m_\ell$, in which case the mass-dependent term in Eq.~\eqref{eq:photonphoton} cancels the anomalous contribution, so that $g_{a\g\g}^{\rm loop} \approx 0$ is a reasonable approximation in that kinematical regime, see Eq.~(\ref{suppressed-gphoton}).

It follows from Eq.~(\ref{eq:photonphoton}) and Eqs.~(\ref{eq:photonZ_anomaly})-(\ref{eq:WW-anomaly}) that the experimental bounds on the ALP-photon coupling $g_{a\g \g}$ lead to bounds on $\Tr(\Cee)$ only, while the experimental bounds on the ALP-$WW$ coupling $g_{a WW}$  can be directly translated into bounds on $\Tr(\Cnn)$. In contrast, the data on the two ALP couplings involving the $Z$ boson  --Eqs.~(\ref{eq:photonZ_anomaly})  and (\ref{eq:ZZ_anomaly})- constrain two (different) combinations of both traces, which define  flat directions of the EFT. In order to avoid such cancellations altogether, simultaneous bounds on at least two different phenomenological ALP couplings  among the three in Eqs.~(\ref{eq:photonZ_anomaly})-(\ref{eq:WW-anomaly}) have to be considered.

\begin{figure}[t]
\includegraphics[width=0.95\textwidth]{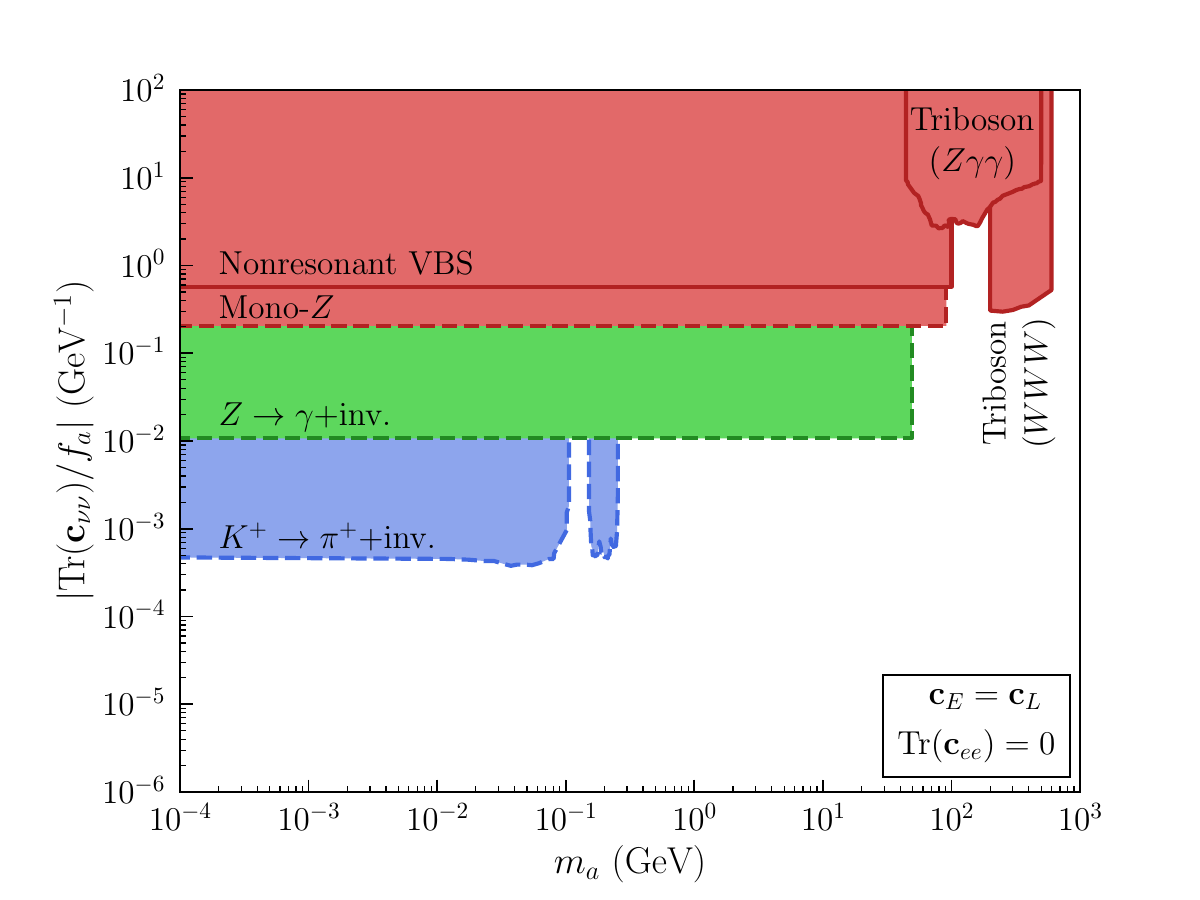}
\caption{Direct and loop induced bounds on $|\Tr(\Cnn)/f_a|$ derived from several experimental searches assuming no ALP-electron couplings: $\Cee = 0$.}
\label{fig.BoundsNeutrinoCoupling_no_cee}
\end{figure}

\begin{figure}[t]
\includegraphics[width=0.95\textwidth]{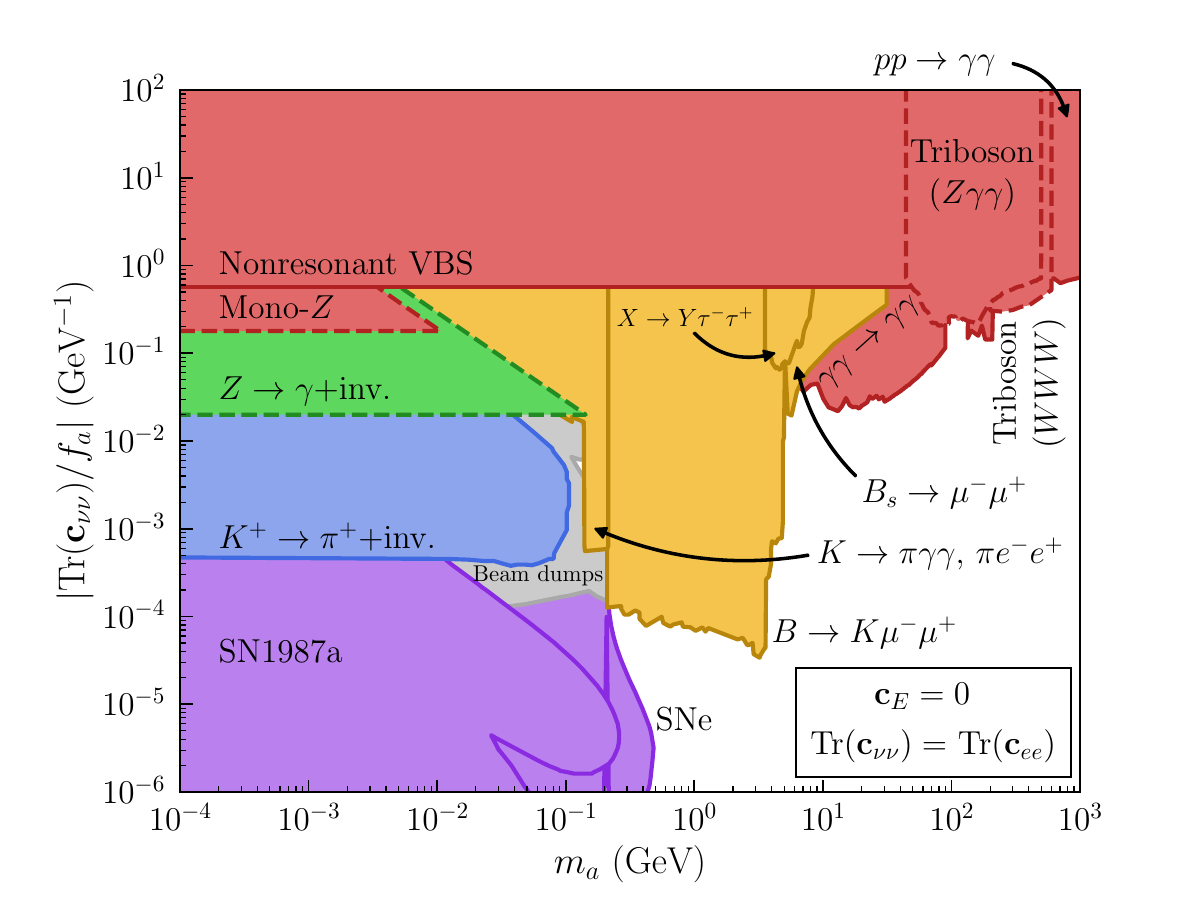}
\caption{Direct and loop induced bounds on $|\Tr(\Cnn)/f_a|$ derived from several experimental searches assuming no ALP coupling to right-handed electrons ($\ce = 0$) so $\Cee = \Cnn = - \cL$.}
\label{fig.BoundsNeutrinoCoupling_all}
\end{figure}

\begin{figure}[t]
\includegraphics[width=0.95\textwidth]{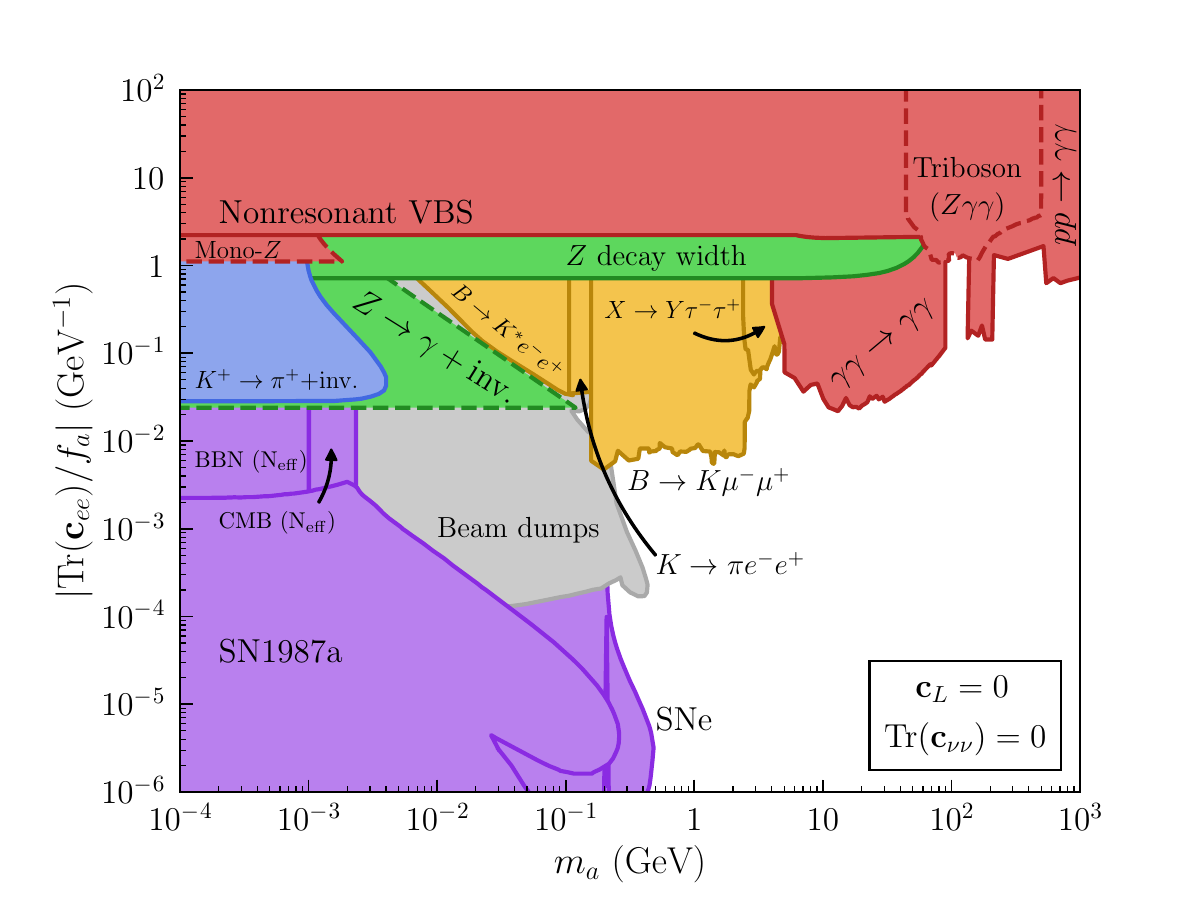}
\caption{Direct and loop induced bounds on $|\Tr(\Cee)/f_a|$ derived from several experimental searches assuming no ALP-neutrino couplings: $\Cnn = 0$.}
\label{fig.BoundsNeutrinoCoupling_no_cnunu}
\end{figure}

 \begin{figure}[t]
 \centering
\includegraphics[width=1\textwidth]{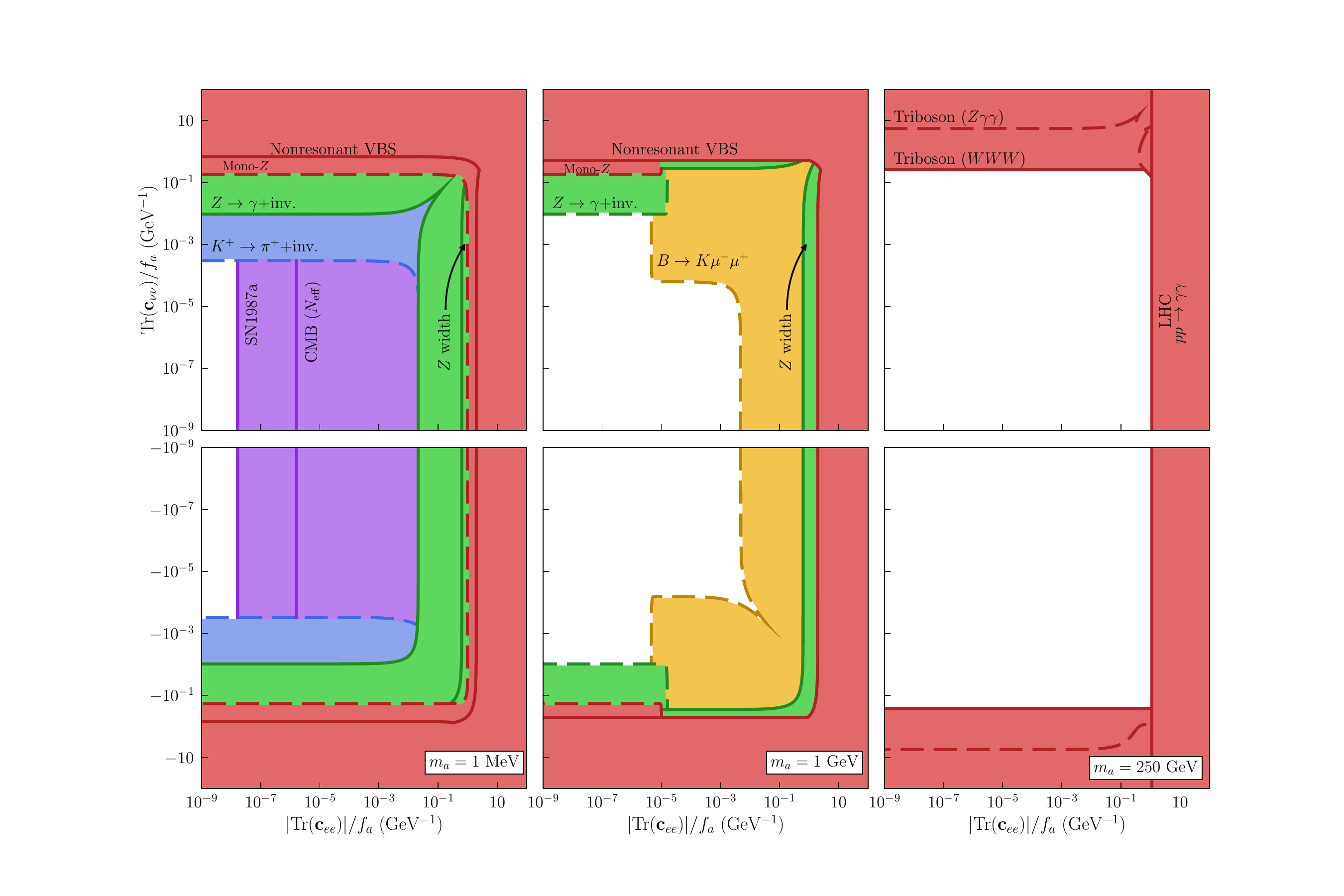}
\caption{Direct and loop induced bounds on the $\left\{|\Tr(\Cee)/f_a|,\, \Tr(\Cnn)/f_a \right\}$ parameter space for different ALP masses. Rare meson decay bounds in yellow and astrophysical bounds in purple have been obtained assuming universal $\Cee$ couplings.}
\label{fig.Bounds_cnunu_vs_cee}
\end{figure}

The  bounds on $\Tr(\Cnn)$ and $\Tr(\Cee)$ relevant for the kinematical region under discussion are illustrated in Figs.~\ref{fig.BoundsNeutrinoCoupling_no_cee}, \ref{fig.BoundsNeutrinoCoupling_all}, \ref{fig.BoundsNeutrinoCoupling_no_cnunu} and \ref{fig.Bounds_cnunu_vs_cee}. They depict the constraints derived from colliders (green and red), rare meson decays (blue and yellow), beam dump experiments (grey) and astrophysics/cosmology (purple). Figs.~\ref{fig.BoundsNeutrinoCoupling_no_cee} and~\ref{fig.BoundsNeutrinoCoupling_all} summarize  
the bounds obtained on $\Tr(\Cnn) / f_a$ as a function of $m_a$, for two distinct values of the ALP coupling to charged leptons $\Cee$. For reference, Fig.~\ref{fig.BoundsNeutrinoCoupling_no_cnunu} depicts instead the constraints on $\Tr(\Cee)/f_a$ assuming $\Cnn=0$. Finally, Fig.~\ref{fig.Bounds_cnunu_vs_cee} sheds a different perspective: it depicts our results in the  $\{\Cnn, \Cee\}$  trace parameter space, for different values of $m_a$. Details on all those figures follow:
\begin{itemize}
\item The {\it photophobic} ALP is depicted in Fig.~\ref{fig.BoundsNeutrinoCoupling_no_cee}. It assumes $(\Cee)_{ii} = 0$, i.e.  no tree-level flavour-diagonal ALP-charged lepton interactions. This holds for $\ce = \cL$, that is, if  the coupling matrices for left-handed and right-handed leptons coincide.  This forbids ALP decays into same-flavour charged leptons (at tree-level) and into photons (at one-loop-level), from which the popular denomination as ``photophobic'' follows~\cite{Craig:2018kne}. Such a benchmark is particularly relevant for invisible searches, in which the ALP is measured as MET.  Indeed, for ALPs lighter than the massive EW gauge bosons, the only possible decay is then $a\to \overline{\nu}\nu$, which is naturally suppressed by neutrino masses. It follows that the ALP can be in practice considered stable at collider distances for masses up to $m_a \leq M_Z$.
 
\item  Fig.~\ref{fig.BoundsNeutrinoCoupling_all} illustrates another interesting -- and complementary-- case: an ALP which couples  to charged leptons and to neutrinos with the same strength, i.e. $\Cee = \Cnn = - \cL$. In other words, it only interacts with left-handed  lepton doublets, i.e.  $\cL\ne 0, \ce = 0$. The limits in the photophobic case --Fig.~\ref{fig.BoundsNeutrinoCoupling_no_cee}-- are approximately maintained, but for a reduction in the LEP and mono-$Z$ sectors  and the strengthening of the constraints from $Z\gamma\gamma$ searches, 
to be discussed below. More importantly,  Fig.~\ref{fig.BoundsNeutrinoCoupling_all} exhibits strong additional constraints from rare meson decays into charged leptons and photons (in yellow), beam dump experiments (in grey) and from astrophysics and cosmology (in purple) given that the ALP-electron channel --and therefore the ALP-photon channel-- is now open. 

\item The bounds obtained on $\Tr(\Cee)$ are presented in Fig. \ref{fig.BoundsNeutrinoCoupling_no_cnunu}. These are relevant to our discussion as they further constrain the ALP-neutrino coupling in those scenarios where both $\Cee$ and $\Cnn$ are non-zero (as in Fig.~\ref{fig.BoundsNeutrinoCoupling_all}).

\item  The parameter space in the  $\{\Tr{(\Cnn)}, \Tr{(\Cee)}\}$ plane is depicted in Fig.~\ref{fig.Bounds_cnunu_vs_cee} for representative values of $m_a$. The white areas in this figure signal the hunting arena available for ALP-neutrino interactions.

\end{itemize}
Previous works in the literature which derived constraints on flavour-diagonal ALP-lepton couplings based on their loop-impact on ALP couplings to EW gauge bosons include:
\begin{itemize}
    \item Ref. \cite{Bauer:2021mvw}, where new bounds on $\ce$ and $\cL$ were derived from rare meson decays data, at the one-loop and two-loop levels, restraining the analysis to one-coupling-at-a-time among  the two ALP-lepton couplings, $\ce$ or $\cL$.
    \item Refs. \cite{Caputo:2021rux,Ferreira:2022xlw} again only consider one of the two couplings. There, new bounds on the physical ALP coupling to muons and electrons, respectively, are derived from their one-loop contribution to the ALP-photon coupling, $g_{a\gamma\gamma}^\text{loop}$, while the ALP-neutrino coupling is disregarded. 
\end{itemize}
In our work we extend the analysis beyond the one-at-a-time paradigm, exploring the bi-dimensional ALP-lepton parameter space, and furthermore novel bounds are extracted from data sensitive to anomalous ALP couplings to $Z$ and $W$ bosons. Additionally, constraints represented in Figs.~\ref{fig.BoundsNeutrinoCoupling_no_cee}-\ref{fig.Bounds_cnunu_vs_cee} as dashed contours signal limits which require extra elaboration or assumptions beyond the direct application of the original references, to be detailed below.

\subsection{Collider searches}

For heavy ALPs with masses  $m_a \gtrsim \mathcal{O} (10)$ GeV, the experimental bounds on their coupling to neutrinos  are dominated by the high-energy  collider searches for beyond the Standard Model (BSM) physics, here depicted in red and green. These include bounds from $Z$ observables at LEP, LHC searches on mono-$Z$ events~\cite{Brivio:2017ije}, searches for non-resonant ALPs in Vector Boson Scattering (VBS) processes~\cite{Bonilla:2022pxu},  light-by-light ($\g\g \to \g\g$) scattering measured in Pb-Pb collisions~\cite{CMS:2018erd,ATLAS:2020hii}, diphoton production at LHC~\cite{Mariotti:2017vtv} and resonant triboson searches for either $WWW$ events~\cite{CMS:2019mpq} or $Z\g\g$ events~\cite{Craig:2018kne} mediated by heavy ALPs. Note that for the non-resonant ALP searches the theoretical expressions in the previous sections apply simply replacing $m_a^2$ by $p_a^2$, where $p_a$ is the ALP momentum. Additional collider bounds derived in scenarios where lepton couplings are set to zero (e.g. refs.~\cite{Aiko:2023trb,Aiko:2023nox}) are not included here, as they do not apply to our analysis.

{\it A remarkable fact will be shown:  in the heavy ALP parameter space region, sizable ALP EW interactions are still allowed at present, which evidences the need for experimental efforts targeting such couplings.}

\begin{figure}[t]\centering
	\includegraphics[width=.24\textwidth]{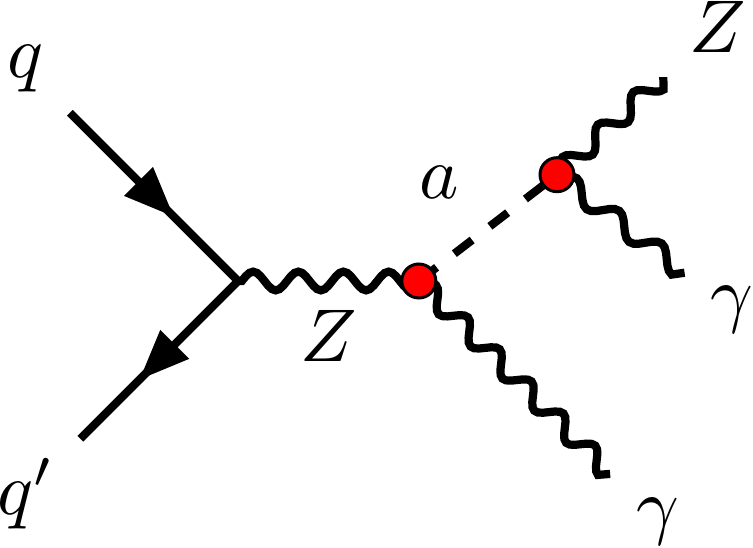}
	\hfill
	\includegraphics[width=.24\textwidth]{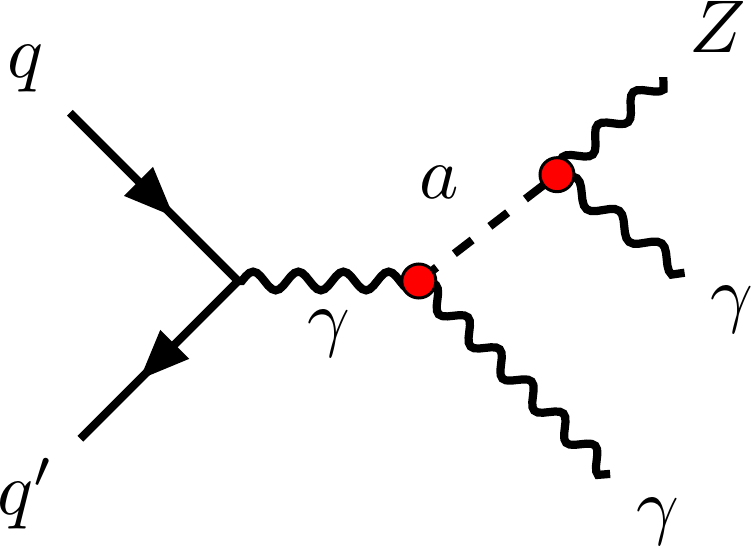}
	\hfill
	\includegraphics[width=.24\textwidth]{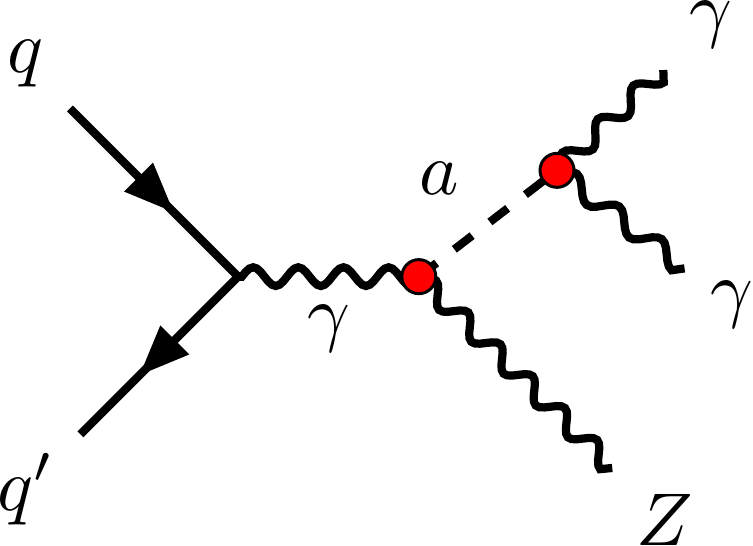}
	\hfill
	\includegraphics[width=.24\textwidth]{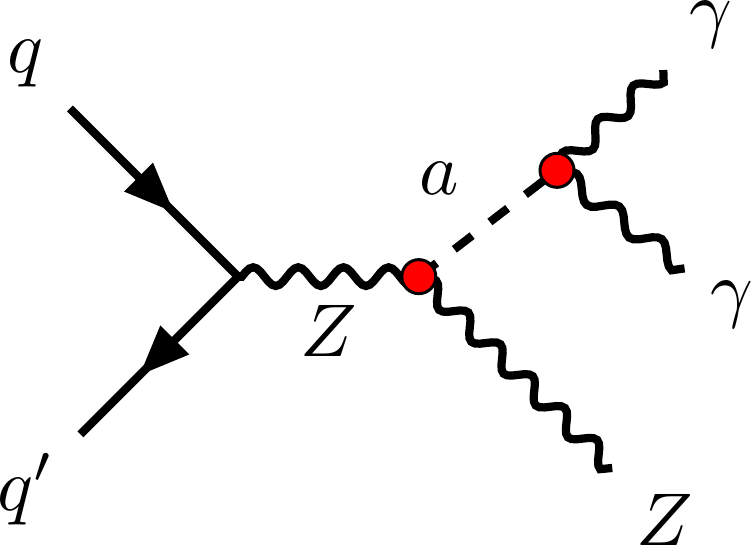}
	\caption{Tree-level diagrams contributing to the ALP-mediated triboson process $pp \rightarrow Z\g\g$. Additional topologies (i.e. vector boson scattering) are suppressed by the kinematic cuts imposed on the original analysis.}
	\label{fig.triboson-diagrams}
\end{figure} 
\subsubsection*{Triboson bounds}
For the triboson data,  $a\rightarrow	W^+ W^-$ is the dominant ALP decay mode when this channel is kinematically open, in both Figs.~\ref{fig.BoundsNeutrinoCoupling_no_cee} and~\ref{fig.BoundsNeutrinoCoupling_all}; this is consistent with the requirement assumed in data analysis~\cite{CMS:2019mpq}. Since at leading order $g_{aWW}^{\rm loop}$  is independent of the ALP-charged lepton coupling $\Cee$ --see Eq.~(\ref{eq:WW-anomaly})-- the small differences in the $WWW$ triboson bound between both figures are only due to mild changes in the branching ratios for ALP decay into $W$ bosons. 

The analysis of the neutral triboson channel $Z\g\g$ is more subtle. The figures show that the $Z\g\g$ triboson bound becomes stronger for $\Cee = \Cnn$ than in the photophobic case ($\Cee = 0$) in Ref.~\cite{Craig:2018kne}, where it was originally computed.  The point is that the five diagrams in Fig. \ref{fig.triboson-diagrams} contribute in the general case with both types of couplings active, while for $\Cee = 0$ the ALP coupling to photons vanishes and only diagram 1 can contribute. The dashed $Z\g\g$ bounds in  Figs.~\ref{fig.BoundsNeutrinoCoupling_all} and \ref{fig.BoundsNeutrinoCoupling_no_cnunu} are qualitative estimations obtained from the photophobic limit depicted in Fig.~\ref{fig.BoundsNeutrinoCoupling_no_cee} via a scale factor computed from the \textsc{MadGraph5\_aMC@NLO} \cite{Alwall:2014hca,Frederix:2018nkq} cross sections of the ALP-mediated process.\footnote{As the ALP-mediated $pp \rightarrow Z\g\g$ cross section  presents a quadratic dependence on the ALP-couplings in the resonant scenario (see Fig. \ref{fig.triboson-diagrams}), we scaled the photophobic bound by a factor $\left(\sigma_\text{ph} / \sigma \right)^{1/2}$, where $\sigma_\text{ph}$ denotes the cross section in the photophobic case, and $\sigma$ that in the alternative scenario.} Overall, the bounds from triboson searches are shown to be superseded by those from non-resonant searches except for a narrow window in  ALP masses.

\subsubsection*{LEP and mono-\tpdf{\boldmath $Z$ \unboldmath}  bbounds}

The bounds from $Z$ observables at LEP are very relevant, in particular those from the $Z$ partial decay width to $\g +$ inv.~\cite{L3:1997exg, Craig:2018kne} and from its total decay width\footnote{These are superseeded in Fig.~\ref{fig.BoundsNeutrinoCoupling_all} by those stemming from  non-resonant VBS limits.}~\cite{Brivio:2017ije,Craig:2018kne}, here illustrated in green. They stem from 
 experimental bounds on $g_{a\g Z}$ and $g_{aZZ}$, which are recasted via the one-loop analysis as limits on combinations of $\Tr{(\Cnn)}$ and $\Tr(\Cee)$ --Eqs.~\eqref{eq:photonZ_anomaly}-\eqref{eq:ZZ_anomaly}--  and thus the extraction of limits on $\Tr{(\Cnn)}$  depends explicitly on $\Tr(\Cee)$.  

The original mono-$Z$ constraint is derived in Ref.~\cite{Brivio:2017ije} for a photophobic ALP, which is shown here in Fig.~\ref{fig.BoundsNeutrinoCoupling_no_cee}. Thus, in order to relate the former with the bound in Fig.~\ref{fig.BoundsNeutrinoCoupling_all}, we  apply a scale factor on the $pp \rightarrow Z a$ cross section using again \textsc{MadGraph5\_aMC@NLO} \cite{Alwall:2014hca,Frederix:2018nkq} as  discussed above.\footnote{Since the ALP-mediated $pp \rightarrow Z a$ cross sections depend quadratically on the ALP couplings, such scale factor is now obtained as $\left(\sigma_\text{ph.ph.}/ \sigma \right)^{1/2}$.} Furthermore, our bounds  from $Z\rightarrow \g + \mathrm{inv.}$ and mono-$Z$ data in  Fig.~\ref{fig.BoundsNeutrinoCoupling_no_cee} extend to ALP masses $m_a > 3 m_{\pi^0}$, as  the hadronization of the ALP is 2-loop suppressed in this photophobic scenario. Finally, to ensure that the ALP remains invisible within the $\Cee \neq 0$ scenario (see Fig.~(\ref{fig.BoundsNeutrinoCoupling_all})) we impose a decay distance larger than 1 m and 10 m \cite{Craig:2018kne, Brivio:2017ije} for the mono-$Z$ and $Z\rightarrow \g + \mathrm{inv.}$ bounds, respectively. 

\subsection{Rare meson decays}
Rare $K$ and $B$ decays provide limits in certain regions of  parameter space, here  represented  in blue and yellow.  The ALP will either escape detection as \emph{missing transverse energy} (MET)  (in blue), or --in the non-photophobic case-- decay next into lighter particles such as charged leptons or photons (in yellow). All these rare processes involve flavour-changing neutral currents (FCNC), which in the SM only happen at loop level and are mediated by $W$ bosons. Thus,  they typically set strong bounds on $g_{aWW}^{\rm loop}$, which can be directly translated into a bound on $\Tr(\Cnn) /f_a$ via Eq.~\eqref{eq:WW-anomaly}. Additionally, in the $\Cnn = 0$ scenario such bounds can be recasted into (weaker) bounds on $\Tr (\Cee)/f_a$ based on its loop-impact on $\cB$ from Eq.~\eqref{eq:EW_couplings} (through the anomalous contribution to the $\g\g$, $\g Z$ and $ZZ$ couplings). The bounds analyzed in this subsection can be inferred from the constraints on $\cL$ and $\ce$ from Figs.~26 and~27 of  Ref.~\cite{Bauer:2021mvw}.  It follows that:
\begin{itemize}

\item
For light ALPs, the best bounds stem from the search for $K^+ \to \pi^+ \overline{\nu} \nu$ decays at the NA62 experiment~\cite{NA62:2021zjw}, i.e. the search for $K^+ \to \pi^+  + \text{invisible}$; this MET analysis can be directly reinterpreted in terms of ALPs  which are stable within the detector. We have adapted those limits~\cite{NA62:2021zjw, Bauer:2021mvw} assuming that the effect of $\mathbf{c}_E$ will be subleading with respect to that arising from $\mathbf{c}_L$.\footnote{This is justified as the bounds obtained in Fig. 27 Ref. \cite{Bauer:2021mvw} in the $\mathbf{c}_E \neq 0$ scenario are two orders of magnitude weaker than those for the $\mathbf{c}_L \neq 0$ scenario.} 

\item For intermediate ALP masses and the general case $\Cee\neq 0$,  Figs.~\ref{fig.BoundsNeutrinoCoupling_all}-\ref{fig.Bounds_cnunu_vs_cee} depict in yellow the limits for  ALPs unstable within collider distances. These includes several searches which are sensitive to: i) the decay rate of kaons into pions plus an ALP that later decays into a $e^+ e^-$ or a $\g\g$ pair (labeled as $K \to \pi \g\g ,\, \pi e^-e^+  $)~\cite{E949:2005qiy,NA62:2014ybm,NA48:2002xke,KTeV:2008nqz}; ii) LHCb searches for resonant ALPs in $B \to K^{(*)} \mu^-\mu^+$ decays~\cite{LHCb:2015nkv,LHCb:2016awg}; iii) the measurement of the partial decay width of $B_s$ into muons~\cite{Albrecht:2019zul}; iv) several decay processes which can be mediated by $a \to \tau^-\tau^+$ ($B\to K \tau^-\tau^+$ and $\Upsilon \to \g a(\tau\tau)$~\cite{BaBar:2012sau}),  represented here generically under the label $X \to Y \tau^- \tau^+$; and v) LHCb measurements of the $B \to K^* e^- e^+$ branching ratio~\cite{LHCb:2015ycz}. Notice that all these searches require the ALP to decay into visible particles within collider distances, and in consequence they do not apply in the scenario $\Cee = 0$. For instance, a maximum decay length of 60 cm (see Ref.~\cite{LHCb:2016awg}) is imposed in Fig.~\ref{fig.Bounds_cnunu_vs_cee} for bounds stemming from $B \to K^{(*)} \mu^-\mu^+$ searches at LHCb. For illustrative purposes, the bounds shown assume universality, this is, equal diagonal matrix elements  $(\Cee)_{ii}$.
 
The channels with final state charged leptons require the tree-level insertion of  ALP-charged lepton couplings $\Cee$ in addition to their one-loop insertion, unlike all other data considered here. 
 
 \end{itemize}

 It is worth mentioning that, in some of the rare meson processes analysed, the ALP and/or some gauge boson(s) $V$ are off-shell, while our expressions above for $g_{aVV'}^{\rm loop}$ were computed for on-shell external particles. Nevertheless,  on one side the generalization to off-shell ALPs is straightforward as explained, and, on the other, the  $g_{aWW}^{\rm loop}$ dependence on momenta is only present in mass-dependent terms which are subleading with respect to the anomalous -- and thus momentum-independent-- contributions, as discussed earlier as well, and thus the bounds depicted are solid in this respect.

\subsection{Bounds on ALP-photon and electron couplings}

Some of the bounds included in Fig.~\ref{fig.BoundsNeutrinoCoupling_all} arise from observables involving the ALP coupling to charged leptons, $\Cee$, either at tree-level or at one-loop level from its impact on $g_{a\g\g}^{\rm loop}$ --see Eq.~\ref{eq:photonphoton}-- independently of the value of the physical ALP-neutrino coupling. However, those bounds still impose limits on $\Cnn$ due to its relation with $\Cee$ via gauge invariance (Eq.~\ref{eq:pheno-couplings}). These include:
\begin{itemize}
    \item \textbf{Astrophysics and cosmology bounds.} For a general non-photophobic ALP, its putative decay into di-photon pairs, Primakoff emission, and ALP production via photon coalescence in a hot and dense environment would contribute as an additional cooling channel for SN1987a~\cite{Ferreira:2022xlw, Diamond:2023scc} --potentially shortening its neutrino burst and producing observable gamma-rays after the explosion,-- and would affect the luminosity evolution of low-energy supernovae \cite{Caputo:2022mah} (labeled as SNe)\footnote{The narrow gap in the SNe bound, corresponding to an ALP mass $m_a = 2 m_\mu$, can be understood from Eq.~\eqref{eq:photonphoton} as a partial cancelation between the electron and muon contribution to $g_{a\gamma\gamma}^{\rm loop}$ at 1-loop.}. Similarly, it would leave a clear signature in the CMB and BBN observations through the impact on the effective number of early universe species for light enough ALPs~\cite{Ghosh:2020vti,Depta:2020zbh}. 
    \item \textbf{Beam dump searches} for ALP-photon and ALP-electron\footnote{We assume the universality of $\Cee$, i.e. diagonal and equally-coupled to all generations, in order to extend these bounds to $\Tr(\Cee)$.} couplings~\cite{CHARM:1985anb,Riordan:1987aw,Blumlein:1990ay,Dolan:2017osp,NA64:2020qwq,Waites:2022tov}.
    \item \textbf{High-energy collider searches} for ALP-mediated diphoton production at the LHC~\cite{Mariotti:2017vtv},  and light-by-light ($\g\g \to \g\g$) scattering in Pb-Pb collisions~\cite{CMS:2018erd,ATLAS:2020hii}.
\end{itemize}
The remaining constraints on $\Cee$ in Fig. \ref{fig.BoundsNeutrinoCoupling_no_cnunu} are derived assuming $\Tr(\Cnn)=0$ (i.e. $\cL = 0$).

\section{Conclusions}\label{sec:conclusions}

The dark sector of the Universe deserves utmost attention as it heralds new laws of physics beyond the Standard Model of Particle Physics. Neutrinos arguably constitute excellent portals in this search, as their interactions are not obscured by strong or electromagnetic forces. We explored here the possible effective interactions of neutrinos to axion-like particles. In addition to theoretical developments, we have derived new bounds on ALP-neutrino interactions.

The model-independent tool of the ALP EFT has been used in order to formulate the problem.  Its inherent gauge invariant nature ties necessarily together the exploration of ALP couplings to neutrinos with those  to charged leptons. Our analysis has thus encompassed both types of couplings,  i.e. the set of $\{\Cee, \Cnn\}$ matrices, exploring how the limits on ALP-neutrino interactions may vary depending on the assumptions about ALP interactions with charged leptons. This is in contrast with previous literature, where typically just about half of that parameter space was tackled.
 
 Neglecting neutrino masses, the trace of ALP-neutrino couplings can be entirely traded by anomalous ALP couplings to the set of EW gauge bosons  $ \{g_{aW W} , g_{aZZ} , g_{a\g Z} \}$ as  they are connected via the  chiral anomaly.  We have explored theoretically the one-loop impact of the former on the latter (in addition to the well known impact of  ALP-charged lepton couplings on them and on $g_{a\g\g}$).  The complete one-loop computations were presented for ALPs and EW gauge bosons on-shell.

We have next recasted existing bounds on anomalous couplings of EW bosons to ALPs in terms of the $\{\Cee, \Cnn\}$ parameter space, confronting data from different type of experiments. These bounds hold barring extreme fine-tuned cancellations between the contributions explored here and other sources of ALP- EW gauge boson anomalous couplings. While similar constraints have been set previously from rare meson decays~\cite{Bauer:2021mvw}, the limits on ALP-neutrino interactions extracted from collider data  are presented here  for the first time. These span extensive novel territory in parameter space for a wide range of ALP masses, ranging from ultralight ALPs to  $\sim 10^3$ GeV ALP masses, and they sweep over several orders of magnitude in strength; data from heavy ALP searches at colliders have a particularly strong impact, see Figs.~\ref{fig.BoundsNeutrinoCoupling_no_cee}-\ref{fig.Bounds_cnunu_vs_cee}, which demonstrate the potential of high energy probes as a tool to explore  ALP EW interactions and thus, indirectly, ALP-neutrino couplings.

ALP-neutrino interactions are being increasingly considered in the arena of neutrino physics, because of their intrinsic interest and for their putative impact on the cosmological history and on astrophysics issues. Hopefully this work will help to clarify the formulation of the problem and to delimitate the frontiers of this quest.

\section*{Acknowledgements}

We acknowledge early discussions with Carlos Arg{\"u}elles, Jorge Fern\'andez de Troc\'oniz and Iv\' an Mart\' inez Soler. We thank Ricardo Zambujal Ferreira and Edoardo Vitagliano for comments on astrophysical bounds. The work of J.B. was supported by the Spanish MICIU through the National Program FPU (grant number FPU18/03047). The work of J.M.R. was supported by the Spanish MICIU through the National Program FPI-Severo Ochoa (grant number PRE2019-089233). The authors acknowledge partial financial support by the Spanish MINECO through the Centro de excelencia Severo Ochoa Program under grant SEV-2016-0597, by the Spanish ``Agencia Estatal de Investigac\'ion'' (AEI), the EU Horizon 2020 research and innovation programme under the Marie Sklodowska-Curie grant agreement No 860881-HIDDeN, and under the Marie Sklodowska-Curie Staff Exchange grant agreement No 101086085-ASYMMETRY, and the EU ``Fondo Europeo de Desarrollo Regional'' (FEDER) through the project PID2019-108892RB-I00/AEI/10.13039/501100011033. 

\newpage

\appendix


\section{Complete expressions for one-loop couplings}\label{app:full-expressions}
We present the results for the one-loop amplitudes associated to the triangle diagram in Fig.~\ref{figdiagram} for non-vanishing neutrino masses $m_{\nu_\ell}$. Only the $g_{aZZ}$ and $g_{aWW}$ anomalous couplings receive additional contributions from $m_{\nu_\ell}$, while $g_{a\g\g}$ and $g_{a\g Z}$ --Eqs.~\eqref{eq:photonphoton} and \eqref{eq:photonZ}, respectively-- remain unchanged. The expressions for $g_{aZZ}^{\rm loop}$ and $g_{aWW}^{\rm loop}$ read
\begin{equation} \label{eq:gaZZ-massive-neutrino}
    \begin{aligned}
        g_{aZZ}^{\rm loop} =  \, - \frac{\alpha_{\rm em}}{2 c_w^2 s_w^2 \pi f_a} \Bigg\{ & ( 1 - 2 s_w^2 ) \Tr (\Cnn)  + 2 s_w^4 \Tr (\Cee)  \\
        & - \sum_\ell \frac{(\Cee)_{\ell\ell} m_{\ell}^2}{m_a^2 - 4 M_Z^2} \Bigg[ \mathcal{B} (m_a^2, m_{\ell}, m_{\ell}) - \mathcal{B} (M_Z^2, m_{\ell}, m_{\ell}) \\
        & + \Big( (1 - 4 s_w^2)^2 M_Z^2  + 2 s_w^2 (1 - 2 s_w^2) m_a^2  \Big) \mathcal{C}_0 \left( m_a^2, M_Z^2, M_Z^2, m_{\ell}, m_{\ell} , m_{\ell}\right) \Bigg] \\
        & - \sum_\ell \frac{(\Cnn)_{\ell\ell} m_{\nu_\ell}^2}{m_a^2 - 4 M_Z^2} \Bigg[ \mathcal{B} (m_a^2, m_{\nu_\ell}, m_{\nu_\ell}) - \mathcal{B} (M_Z^2, m_{\nu_\ell}, m_{\nu_\ell}) \\
        & + M_Z^2 \mathcal{C}_0 \left( m_a^2, M_Z^2, M_Z^2, m_{\nu_\ell}, m_{\nu_\ell} , m_{\nu_\ell}\right) \Bigg] \Bigg\}\,.
    \end{aligned}
\end{equation}

    \begin{equation}
    \begin{aligned}
    \label{eq.gaWW-massive-neutrino}
    g_{aWW}^\text{eff} = - \frac{\alpha_{em}}{2 \sw^2 \pi f_a }\Bigg\{ & \Tr{(\Cnn)} - \sum_\ell\frac{(\Cee)_{\ell\ell} m_\ell^2}{m_a^2 - 4M_W^2}\Bigg[\mathcal{B}(m_a^2,m_\ell,m_\ell) - \mathcal{B}(M_W^2,m_\ell,m_{\nu_\ell}) 
    \\ & -\left( 1- \frac{m_\ell^2}{M_W^2} \right) \left( \log \left( \frac{m_\ell}{m_{\nu_\ell}} \right) - M_W^2 \mathcal{C}_0 (m_a^2,M_W^2,M_W^2,m_\ell,m_\ell,m_{\nu_\ell}) \right) \Bigg] 
    \\ & - \sum_\ell\frac{(\Cnn)_{\ell\ell} m_{\nu_\ell}^2}{m_a^2 - 4M_W^2}\Bigg[\mathcal{B}(m_a^2,m_{\nu_\ell},m_{\nu_\ell}) - \mathcal{B}(M_W^2,m_\ell,m_{\nu_\ell}) 
    \\ &+\left( 1 - \frac{m_{\nu_\ell}^2}{M_W^2}\right) \left( \log \left( \frac{m_\ell}{m_{\nu_\ell}} \right) + M_W^2 \mathcal{C}_0 (m_a^2,M_W^2,M_W^2,m_{\nu_\ell},m_{\nu_\ell},m_\ell) \right) \Bigg] 
    \\ & + \sum_\ell\frac{ m_\ell^2 m_{\nu_\ell}^2}{M_W^2 (m_a^2 - 4M_W^2)}\Bigg[ (\Cee - \Cnn)_{\ell\ell}\log \left( \frac{m_\ell}{m_{\nu_\ell}} \right)
    \\ & -(\Cee)_{\ell\ell}M_W^2 \mathcal{C}_0 (m_a^2,M_W^2,M_W^2,m_\ell,m_\ell,m_{\nu_\ell}) 
    \\ & -(\Cnn)_{\ell\ell}M_W^2 \mathcal{C}_0 (m_a^2,M_W^2,M_W^2,m_{\nu_\ell},m_{\nu_\ell},m_\ell) \Bigg] \Bigg\}  \,.
    \end{aligned}
    \end{equation}
    
Note that  these expression are also valid in the regime for off-shell ALPs upon the replacement $m_a^2 \to p_a^2$.

\footnotesize

\bibliography{bibliography}{}
\bibliographystyle{BiblioStyle}

\end{document}